# From England to Italy: the intriguing story of *Poli's engine* for the King of Naples

Salvatore Esposito
Istituto Nazionale di Fisica Nucleare, Naples' Unit
Complesso Universitario di Monte S. Angelo
Via Cinthia – 80126 Naples, Italy
salvatore.esposito@na.infn.it

*Abstract*: An interesting, yet unknown, episode concerning the effective permeation of the scientific revolution in the XVIII century Kingdom of Naples (and, more generally, Italy) is recounted. The quite intriguing story of Watt's steam engine prepared for serving a Royal Estate of the King of Naples in Carditello reveals a fascinating piece of the history of that Kingdom, as well as an unknown step in the history of Watt's steam engine, whose final entrepreneurial success for the celebrated Boulton & Watt company was a direct consequence. That story unveils that, contrary to what claimed in the literature, the first introduction in Italy of the most important technological innovation of the XVIII century did not take place with the construction of the first steamship of the Mediterranean Sea, but rather 30 years before that, thanks to the incomparable work of Giuseppe Saverio Poli, a leading scholar and a very influential figure in the Kingdom of Naples. The tragic epilogue of *Poli's engine* testifies for its vanishing in the historical memory.

## 1. Introduction

The introduction in England of the *steam engine* marked unequivocally the history of the industrialization in the long XVIII century, commonly associated primarily (if not exclusively) to coal mining or factory production. The key figure of James Watt easily emerges in this respect, although surrounded by his predecessors and competitors, who specialized in the appropriate technology that was instrumental to the development of the industrial revolution. Of course, as well crucial for such development were entrepreneurial abilities (Roberts 2009, 2011) of people that had to combine technological abilities with an entrepreneurial stance, in order to accomplish the increasing demand for machines and components used in manufacturing, mining and infrastructural projects, as well as to face issues of problem-solving and of technology type. Just these latter abilities, indeed, stimulated further innovation, also enabling to adapt given technologies to other fields of convergence, resulting in steam engine design and construction practices modelled on both material availabilities and informed abilities.

It is quite important to note (Roberts 2000), however, that a significant part of the early history of steam technology had *de facto* not to do with coal mining or factory production, but rather with water drainage, as well as water and land management, along with garden landscaping and milling. Outside Britain, quite illuminating are the French and Dutch cases. The marquis Antoine de Ricouart d'Herouville carried out a drainage project to pump out excess water in the Moors (Dunkerque) by using steam power, while William Blakey's commission of the Duc of Chartres was aimed to construct a steam-powered installation for watering his *folie de Monceau* (Roberts 2011). Also, a competition was issued by the Batavian Society, concerning problems associated with adapting steam to the management of water around Rotterdam (Roberts 2009).

In the present paper we report on a different illuminating example – though unknown – referring to places and people who are often overlooked when studying such matters, that is the Kingdom of Naples and the Bourbon Court.

The Bourbon Kingdom of Naples was certainly among the most important states of the XVIII century's Europe, playing a unique and incomparable role for Italy and the entire Mediterranean area during Enlightenment. Being the southern Italian trading emporium, boasting an advanced banking system, Naples continued to be the third capital in Europe (after Paris and London) also after the disastrous plague that afflicted it in 1656 (before that event, it was the most populous town in Europe). A general renovation of the Kingdom, compared to previous dominations, occurred with the instauration of the Bourbon family, which took place with the advent of Charles III in 1734 (Croce 1992).

A number of important plans were, then, developed and carried out, such as the building of novel royal residences, public works and factories. Such projects required the investment of considerable economic resources, which, in addition to obvious architectural aspects, served to solve also important complementary problems such as, for example, water supply. The architectural culture of the XVIII century, indeed, acknowledged the relevance of hydraulic works, since their practical execution required high scientific and technical skills, and the Bourbons constantly favored, in the second half of the century, the updating of the competencies of technicians involved in the building of aqueducts, bridges, roads and harbor works (Serraglio 2003).

One of the most important works of this epoch was certainly the realization of the *Acquedotto Carolino*, planned to supply with water the *Belvedere di San Leucio*, a Royal Palace near Caserta. The realization of the *Real Sito di Carditello*, another Royal Estate built in the second half of the XVIII century by Ferdinand IV (son of Charles III) near Capua to serve as a farmhouse, is certainly far less famous, but equally important for the significant works of hydraulic engineering realized there. While it is quite known this additional testimony of the architectural experimentation carried out by the technicians of the Bourbon House in Naples, practically unknown is instead the interesting (for several reasons) story behind the realization of the water supply service for this Royal Palace. The key man who managed for such realization is as well not known, notwithstanding he was the very author of the introduction in the Kingdom of Naples (and, more in general, in Italy) of the certainly most important technological innovation of that epoch – Watt's *steam engine*. His name is nowadays recalled practically only by natural scientists (and few historians of physics), although Giuseppe Saverio Poli was one of the most influential figures of the Kingdom of Naples between the end of the XVIII and the beginning of the XIX century (Esposito 2020). It was just through the agency of Poli that, as assessed in a previous work (Esposito 2017), the basic roots of the scientific thought as well as technological innovations penetrated effectively (and long-lasting) in the most important Kingdom of Italy (and, as recalled above, one of the most important ones in Europe).

Here we will describe in full detail this historical case, which is not at all known in the literature, although, quite curiously, it mimics in several respects what occurred in the Netherlands around the same time (Roberts 2004). The introduction in Italy of the technological prodigy of the XVIII century – the steam engine – is usually dated to 1818, when the first steamship of the Mediterranean Sea was built (again, in the renamed Kingdom of the Two Sicilies). However, as we will see, the steam engine was introduced in Naples just 30 years *before* that event, and the related story is a fascinating piece of the history of that Kingdom, as well as a relevant piece of the history of Watt's steam engine.

In the following Section, we will give a short account on Giuseppe Saverio Poli, one of the protagonists of our story, pointing out his unique role in introducing the first fruits of the scientific and technological revolutions in the Kingdom of Naples, while in Section 3 we briefly summarize the milestones in the history of the steam engine, with particular reference to the work of Boulton and Watt, as also acknowledged by Poli. The intriguing story of Poli's engine is, instead, fully recounted in the subsequent Sections, from its conception to its successful establishment, and even beyond, since that story did not end with the engine erection. The tragic epilogue, causing even the cancelling of the historical memory of such episode, is finally highlighted in Section 7, along with few other concluding remarks.

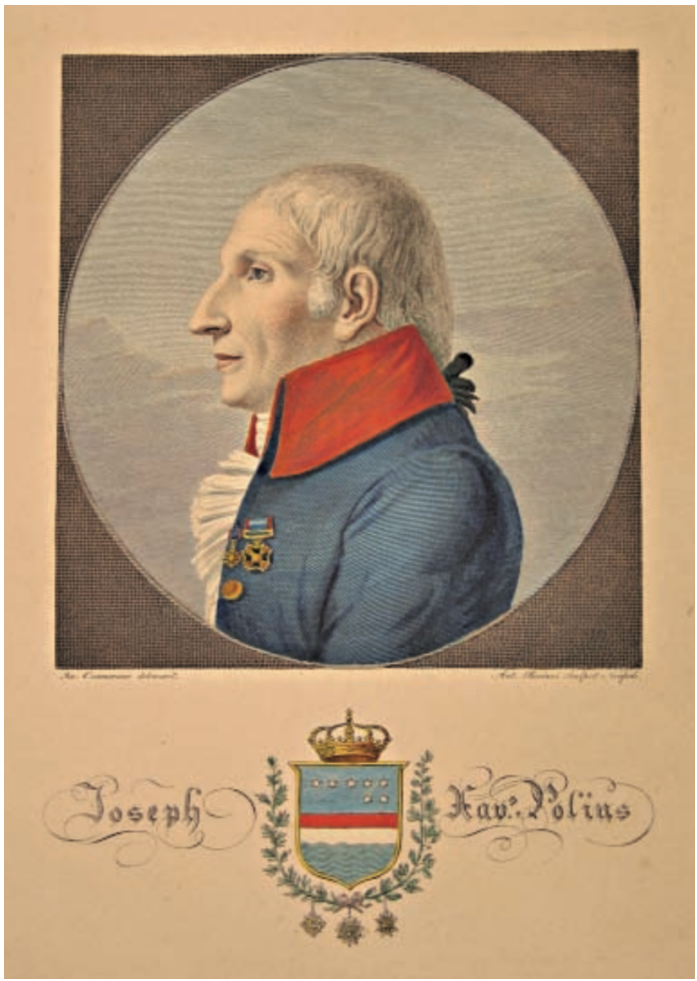

**Fig. 1.** Portrait of Giuseppe Saverio Poli (Poli 1826).

## 2. The perfect scholar and the man of power

Giuseppe Saverio Poli (Fig. 1) was born in the Apulian town of Molfetta on October 28, 1746. After completing his first studies at the local *Seminario Diocesano*, in 1765 Poli was sent to Padua, where he enrolled to study Medicine with L.A.M. Caldani and G.B. Morgagni (and having A. Scarpa as fellow student). He also pursued classical education with J. Facciolati, talking fluently in Greek, Latin and French, and studied physics, astronomy (with Giuseppe Toaldo), botany and natural history. After some stays in Venice, Ravenna, Bologna, Genova, Florence and Rome, he came to Naples in 1771, where he got his Master's degree ("dottorato") in Medicine; he also practiced for some time as a medical doctor (till 1776), but soon he switched his attention to the study of natural sciences. Despite his degree (and even before his accomplishment), indeed, his scientific interests now centered mainly about physics, with particular reference to the current debate about electricism (Esposito 2020).

In 1774 Poli was called, with the rank of Second Lieutenant, to teach Geography and Military History at the Royal Academy of the Corps of Cadets (*Reale Accademia del Battaglione Real Ferdinando*, later *Reale Accademia "Nunziatella"*) created by Ferdinand IV of Bourbon by merging the old artillery academy with that of engineers. Promoted to Lieutenant, Poli was then sent abroad in order to acquire instruments for the scientific cabinet of the Academy. He took this opportunity (1777-1779) to visit the most important academies and institutes in Italy, Germany, France, England and Holland, and study mechanical application methods. In any of these countries Poli received high recognition for his vast erudition, and his valued works on electricism opened to him the doors of the major scientific academies of Europe. In this respect, quite relevant is his election in 1779 as Fellow of the Royal Society of London:

the honor to be a *home member* of such Society was reserved to very few non-British people. Also, during his journey Poli naturally encountered the major physics instrument makers, such as Dollond, Ramsden, etc., with whom he forged close relations not limited to his task to acquire scientific instruments for the Naples' Academy.

When coming back to Naples, in 1780 Poli was appointed to the Chair of Physics at the *Real Collegio Medico-Chirurgico* at the *Regio Ospedale di S. Maria del Popolo* (*Arcispedale degl'Incurabili*) and taught Experimental Physics both at the *Incurabili* and at the *Nunziatella*. The textbook for the course(s) of Experimental Physics, *Elementi di Fisica Sperimentale composti per uso della Regia Università*,[1] was published, in its first edition, in the same year 1781 and testifies (Poli 1781) how, in teaching mechanics and other subjects, Poli adopted all sort of experimental demonstration devices. The extreme clarity of exposition, along with an abundance of examples, made the fortune of this textbook, which counted – from 1781 to 1824 – a total of 23 new editions and reprints (and a posthumous reprint of 1837). It was, no doubt, the most widely used (Italian or foreign) textbook of physics in Italy between the end of the XVIII century and the first quarter of the subsequent one (Esposito 2020).

Notwithstanding the striking success of the *Elementi*, the most important scientific contribution by Poli is its outstanding work *Testacea Utriusque Siciliae eorumque Historia et Anatome tabulis aeneis illustrata*[2] on comparative anatomy and classification of mollusks in the Kingdom of the Two Sicilies. (Poli 1791, 1795, 1826). This work is widely recognized to have effectively laid the foundations of malacology (Temkin 2012) and was acknowledged by some of the major scientists of the XIX century, including Charles Darwin (Esposito 2019). Indeed, it was the first treatise on molluscan biochemistry and physiology, coming from the accurate experimental studies performed by the author, documenting in detail various aspects of shellfish morphology, and reporting a number of novel characters discovered by Poli himself.

After a second "diplomatic" journey to Holland and Germany in the wake of the Duke of Gravina, in 1784 he was chosen by King Ferdinand as tutor to the Crown Prince Francis I for his encyclopedic culture, later being charged of many important court assignments. Since then, an indissoluble link with the Royal family was settled. Due to the events – at the end of 1798 – followed to the French invasion, the Bourbons were forced to take refuge to Palermo, and Poli readily followed the Royal family in their exile. With the first Restoration, however, after the brief period of the Parthenopean Republic (January-June 1799), Poli came back to Naples, and in 1803 he was nominated Commander of the "Nunziatella" Royal Military Academy.

Returning from his second exile in Palermo with the Royal Bourbon Court during the decade (1806-1815) of French domination in the Kingdom of Naples, in 1816 he was elected member (and later, in 1819, became President) of the *Reale Istituto d'Incoraggiamento alle Scienze Naturali* (Royal Institute for the Encouragement to the Natural Sciences). In 1820, following the disorders in the Kingdom aimed at establishing a constitutional monarchy, the King Ferdinand I of the Two Sicilies was forced to grant the Constitution and create a Constitutional Parliament with democratically elected members. However, the King managed to have also trustworthy people in that assembly and, among the others, on November 20 of that year Poli became State Councilor in the Constitutional Parliament. Finally, just one year before his death, Poli was also charged as *Ufiziale alla immediazione di Sua Altezza Reale* Francis I (Official to the immediacy of His Royal Highness), following his beloved Prince in his journeys throughout the Kingdom (Esposito 2019). Poli died on April 7, 1825 – at the age of 78 – in his home in Naples.

## 3. With the eye of the witness: the powerful *steam engine*

During his journey in England, the curious scholar Giuseppe Saverio Poli "discovered" the existence of the marvelous steam pump (named *tromba a vapore* or *tromba a fuoco*) – that is a steam engine-based

---

[1] Elements of Experimental Physics, composed for the use in the Royal University.
[2] Shelled animals of the Two Sicilies with their description and anatomy.

pump – whose description and functioning, along with its possible applications, was readily reported in his lectures on Experimental Physics delivered in Naples, and then incorporated in his textbook *Elementi di Fisica Sperimentale*:[3]

> It is termed steam pump since its working power is due neither to men nor animals, but rather to the steam coming from boiling water. Continually rising from a big boiler full of water placed above a small furnace, this steam penetrates into a pump, where alternatively produces vacuum and plenum. […]

> Its applications and advantages are innumerable, since very great is its efficacy to raise any quantity of water to any height, as well as to provide mills and ship-canals with water, to dry lakes and marshes of any magnitude, to produce uninterrupted and regular motions in any direction […].

Differently from other authors of science textbooks, Poli described here what really saw with his own eyes – a feature that, among the others, likely explains the great success of the *Elementi*. According to his account:

> In the famous manufacture of the ingenious Mr. Boulton in the town of Birmingham there is a large number of devices working by means of such a pump, which also provides the water for a ship-canal. I saw those pumps in other Counties in England used in some foundries, in order to let huge blowing apparatuses working in iron furnaces.

> It is useful to know that the power of such an engine is completely unbounded, since it can be raised – in a sense – to infinity by increasing the proportions of its parts. It works day and night with no interruption and can be stopped at any moment very easily.

The eyewitness Poli did not fail to realize that at the heart of such marvelous, different devices was the *steam engine* improved by the "incomparable Mr. Watt", and the teacher Poli did not miss the opportunity to let his students in Naples to see with their eyes how such a technological prodigy worked, just by means of a model built on purpose on his order:

> The pump to raise water can be divided from what the English terms Steam Engine, consisting just in the cylinder – where we noticed the steam from the boiling water enters – from which the power of the engine develops. Then, by means of just such an engine and without the help of any water but that in the boiler, paper mills, grinding mills and any other device can work, sometimes with an infinite advantage, especially in countries poor of water.

> An excellent model of such an engine, built on my order, can be seen in the rich cabinet of our Royal Military Academy. By means of it, acting as a small grinding wheel, an appreciable quantity of water can be raised in a pump; as well, it can be used to work a mill that effectively grinds wheat, or scutches flax, or even can be used to act some hammers in a rolling mill. It has the advantage to be built with the latest improvements introduced by the incomparable Mr. Watt.

As well known, the Scottish James Watt (Dickinson & Vowles 1948) was not the inventor of what effectively was the moving force of the Industrial Revolution, that is the steam engine – a heat engine performing mechanical work, using steam as its working fluid –, but no doubt that what Watt conceived and constructed was a key component of paramount importance for the development of that Revolution, since, as noted by Poli, it barely allowed factories to locate even where waterpower was not easily available.

Interestingly enough, among the first studies concerning the possibility to exploit the motive force of heat we find just those of Giovanni Battista Della Porta in the same Naples at the beginning of the XVII century (Piccari 2007). The Neapolitan scholar, indeed, realized how the water contained in a tank could be pushed upwards by the steam pressure acting on its surface, the steam condensation being able to produce a suction capable to draw water from a lower level. The intuitions and speculations of such a scholar about the steam elastic force, contained in the *Tre libri de' spiritali* (Della Porta 1606), introduced in the scientific community the appropriate ideas that later allowed the realization of the steam engine, which, not by chance, took place instead in England. The English technical skills for the production of water suction systems using the steam power became, indeed, particularly advanced during the subsequent century, since very frequently emerged there the problem to drain the mines ducts from water infiltration.

---

[3] Here and in the following, see Volume III of (Poli 1794). Translation from the Italian is by the present author.

It was Thomas Newcomen to develop the first successful *atmospheric* steam engine using a piston (Landes 1969), by improving a previously patented (1698) *fire engine* by Thomas Savery. He used it in 1712 to remove flood water from a mine, being essentially designed for pumping: the vacuum created in it was used to suck water from the sump at the bottom of a mine or, more in general, to raise water from below. It was employed mainly for draining mine workings at depths hitherto impossible, as well as for providing reusable water supply to drive waterwheels in factories.

Watt's later paramount contribution was twofold. First of all, it enabled an engine of a given size to do more work and with less fuel than the Newcomen engine, thus improving the power, efficiency and cost-effectiveness of steam engines. Secondly, and even more importantly, it widened the field of application of steam engines to a level that was impossible with the earlier engines. This was accomplished by employing several devices able to convert the reciprocating motion of the piston to a rotary motion suitable for driving factory machinery, such as for grinding, weaving and milling.

Watt's early engines used about half (Hunter & Bryant 1991) as much coal as earlier Newcomen's (as improved by John Smeaton) but, notwithstanding his successful ideas, also directly translatable into workable designs, still a long time had to pass for Watt in order to construct a full-scale engine. The main reason was – obviously – the need of some capital to invest in those ideas, and a partnership was then imperative for him, who continued to work in the meanwhile at perfecting his steam engine with several experiments (Dickinson & Jenkins 1927) which, of course, costed much money too. The true turning point came when the businessman Matthew Boulton began to correspond with Watt (Andrew 2009), by realizing that a steam engine would help him to provide the necessary power for his Soho Manufactory in Birmingham. In 1774 Boulton convinced Watt to move to this place, where an engine was erected and satisfactorily run. The subsequent year Boulton and Watt entered a partnership whose aim was no more limited to produce sufficient power for the Soho Manufactory, but rather aimed at a profitable business venture for efficient steam engine production.

Starting from 1776, the Boulton & Watt firm installed the first engines in commercial enterprises, beginning in Staffordshire and Shropshire (Dickinson & Jenkins 1927). They were typically not manufactured by Boulton & Watt, but rather they purchased parts made by others according to drawings made by Watt, and then assembled on-site. Watt served in the role of consulting engineer, as well as supervisor – at first – in the erection of the engines: these were, indeed, large machines, which required the construction of a dedicated building to house them. Only later, with the expansion of the Soho firm, different engineers or other firm's employers took over engine's erection and on-site management, allowing Boulton and Watt to remain in Birmingham (Andrew 2009).

In the subsequent years, Watt made many improvements and modifications to the steam engine, whose field of application greatly widened when in 1782 he was able (urged by Boulton) to convert the reciprocating motion of the piston to produce a rotary motion, required in mills and factories. Further study and experimentation (especially on rotative beam engines) led Watt to another important technological innovation, by moving from *single-acting* pumping engines to *double-acting* ones, where steam was used to press alternately on the opposite sides of the piston, thus enabling the engine to make a power stroke in both direction (and to not introduce an external force to retract the piston). The project of a *double-acting* engine was effectively brought forward only in 1782-1783 when considering rotative engines, with the further innovation of the so-called *parallel motion* device producing the straight-line motion required for the cylinder rod and pump from the connected rocking beam, whose end moves in a circular arc.

A number of other improvements, modifications and far more experimentation kept Watt busy for still a long time, not only involved in the solution of purely technical problems, but also committed to ensuring a full economic success to the enterprise with Boulton against patent infringements, piracy and marketing problems (Dickinson & Jenkins 1927). As a matter of fact, even though Boulton & Watt erected the first reciprocating, single-acting steam engine in the Continent as early as in 1780, the first rotative engine (outside Soho) in 1783 and the first double-acting engine the following year, it was not before 1788 that Watt was allowed to relax from business, being assured the success of his engine. What specifically made this fact clear is, likely, not very well known in the literature, and remarkably required the agency of an unexpected actor, as we are going to recount in the following section.

### 4. In need of watering the Royal prairies of Carditello

Always anxious to achieve a complete successful enterprise, as a matter of fact Watt realized he had finally reached his goal when one of his engines was ordered from the King of Naples through Giuseppe Saverio Poli's intermediation. As reported in his *Elementi* (Poli 1794):[4]

> [The engine] that I had built in England for the service of His Majesty our Most Clementant Sovereign, and which is already established near the fortresses of Capua to raise the waters of the River Volturno, to be able to water the Royal prairies and the fields of Carditello during summertime, has three feet in diameter, and is able to lift 500 cubic feet of water up to the height of 25 feet in the stretch of each minute; and consequently 30 thousand cubic feet in one hour time. Consider each one as an immense copy that raises it in the interval of 24 hours!

The Royal Estate of Carditello (Fig. 2), near Capua, was practically a farm for the use of Ferdinand IV of Bourbon, King of Naples. Its origins traced back to 1745, when Charles III (the father of Ferdinand) implanted a horse farm in the forests and marshes of the *Real Difesa di Carditello*. As noted by a Neapolitan historian (Giustiniani 1797),

> Ferdinand IV called a not so large population [to Carditello] to attend the herds of cows, buffaloes and horse races, which he wanted there, and also to make good cheeses as in Lodi […]. There are also eight well-built big sheds for the people involved in the job. Finally, there was a forest of oaks, elves, and other wild trees, in which there are wild boars, goats, foxes and hares for King's hunting.

The works for Ferdinand's farm were completed in 1785, but very soon the problem of irrigating the fields of the Royal Estate emerged, and the same Ferdinand IV did not hesitate to assign such a problem to the reliable Court man and scientist Poli, who then immediately thought of contacting an old acquaintance of his in England.

#### 4.1. Poli meets Boulton

As recalled above, Poli visited England at the end of 1770s, especially searching for instrument makers able to provide him with appropriate instruments for the Physics Cabinet of the Royal Military Academy in Naples, including the leading technician Jesse Ramsden (McConnell 2007). According to Boulton, it was just Ramsden that introduced Poli to him,[5] and already in that occasion Poli talked with Boulton about a possible engine to be sent in Naples at the order of the King:

> During my stay in England we [Boulton & Poli] had a long conversation together about a fire Engine, that his Sicilian Majesty would perhaps have settled at Naples. [...] You informed me further that an Engine capable to raise ten thousand cubic feet of water one foot high in one minute's time would cost one thousand Pounds besides the building, as it appears from the note you favoured me with.[6]

---

[4] The news was also reported in various Italian journals starting from 1795, probably taken just from Poli's *Elementi*.
[5] "Mr. Poli is a Gentleman that was first introduced to me by Ramsden who made him a philosophical apparatus at his Sicilian Majesty expenses". Matthew Boulton to James Watt, 26 September 1786. The Library of Birmingham: Birmingham Archives and Heritage. Boulton & Watt Collection: MS 3147/3/10.

Poli visited the Boulton & Fothergill company in Birmingham in November (or October) 1779, accompanied by the Venetian Ambassador Simon Cavalli, who purchased some items.[7] During that occasion, the two Italians had of course the opportunity to see some steam engines in the Soho Manufactory which, however, still were reciprocating single-acting pumping engines able, mainly, to extract water from the mines of the Cornwall. Nevertheless, Poli quickly realized the usefulness of those engines, and Boulton as well did not fail the occasion to promote his products designed by Watt:

> At Birmingham [...] you was so good as to spend an evening there with me, and the Venetian Ambassador [Simon Cavalli]. There you set down all the dimensions, and prices of the Engine at large, with all the least particulars relating to it, so that it takes up a whole sheet of paper.[8]

### 4.2. *Favorable circumstances*

Despite the genuine interest in Watt's steam engine, at the time of Poli's trip to England in 1779-1780 no "favorable circumstances" emerged for the King of Naples to acquire that technological prodigy. The situation changed when the Royal Palace at Carditello was completed. When the effects of the hot summertime of 1786 appeared, Poli promptly wrote to Boulton on behalf of King Ferdinand IV to order a steam engine able to irrigate the fields of the Royal Estate:

> His Majesty being proposed of a meadow of a vast current within 15 miles from Naples; and being desirous to cover it regularly every year since the month of May till that of September, has thought proper to make use of a fire Engine, which might raise upon the water from the River Volturno, and dispose it by the means of several little canals through the meadow abovementioned: it seems to me that all the circumstances are favourable to his scheme, as the water is not to be raised any higher than 27 French feet in order to carry to such a point that is about five French feet above the natural level of the ground which wants watering. The side of the River, which being the nearest to the meadow is the proper place for settling the Engine, is about two English miles distant from the meadow itself. The bed of the River is very large there; and the height of its water during the summertime when the lowest of all, amounts to four English feet. At two miles distance from the place of the Engine there is a large forest, which might afford the necessary quantity of wood.[9]

King's order turned out to be rather compelling and, at the same time, Poli likely misinterpreted Boulton's early promotion or, even, misunderstood the difficulty in making and assembling a powerful steam engine. As a matter of fact, he ingenuously continued his letter to Boulton as follows:

> His Majesty will send two of his Frigates to England, perhaps about the month of September; and they shall remain there for some time. Therefore, orders will be given to their Captains to get you on board the ship the Engine and carry you to Naples, where you may depend to be very graciously received from his Majesty who is well acquainted with your great abilities; so that I am sure you will not repent to have undertaken such a voyage.

---

[6] Joseph Poli to Matthew Boulton, 26 November 1787. The Library of Birmingham: Birmingham Archives and Heritage. Boulton & Watt Collection: MS 3147/3/518/N09.
[7] "L'on m'a remis les petits articles, que avec M.r Poli j'avoir choisi, et je me suis mis en règle avec M.r Mattheus, mais j'aurais souhaité d'y trouver aussi les boutons d'acier, selon l'échantillon, que j'avoir mis de côté, fond uni, avec un contour de clous, et un groupe au milieu, avec du violet, pout habit, veste, et coulotte; et je vous prie de donner vos ordres, afin qu'ils me soient envoyés, et fans le même tems un cabaret de figure ronde, grand pour un service à thé, de vernis violet, quelque figure au milieu, et avec le contour d'argent plated". Simon Cavalli to Matthew Boulton, 29 November 1779 (Zorzanello 1984).
[8] Joseph Poli to Matthew Boulton, 26 November 1787, *cit*.
[9] Joseph Poli to Matthew Boulton, 4 July 1786. The Library of Birmingham: Birmingham Archives and Heritage. Boulton & Watt Collection: MS 3147/3/518/N01.

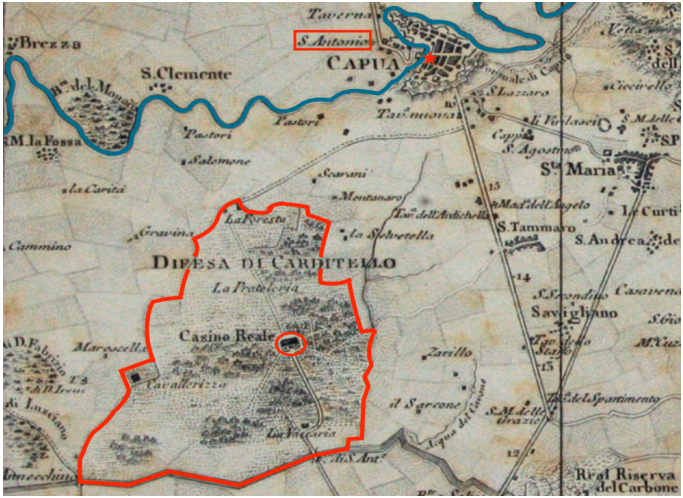

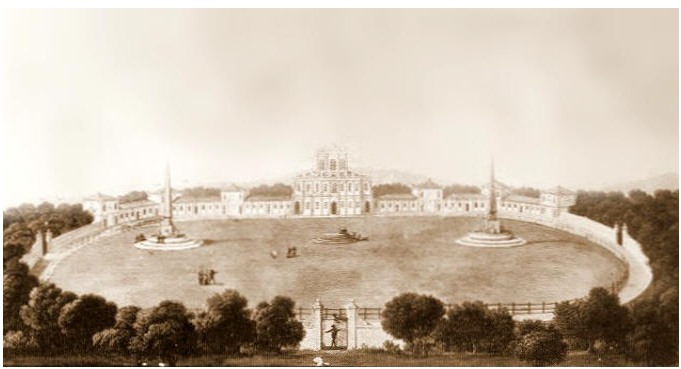

**Fig. 2.** The Royal Estate of Carditello. Top: Map of the *Real Difesa di Carditello* from (Rizzi Zannoni 1812). The house of Poli's Engine was erected on the river Volturno (highlighted in blue) near the town of Capua (*Fortini di Capua*; see the red star), not very far from the place where the *Molini di S. Antonio* were later built. Bottom: a view of the Royal Palace (*Casino Reale*) of Carditello, as from an ancient engraving.

> I will inform you Sir, that his Majesty will reward you for your trouble besides the cost of the Engine, and will grant you a Patent in case you wish to hire it. This, if I am not mistaken, was the particular condition you mentioned me to be assured of before your departure from England. So that I hope nothing will hinder you from getting the Engine ready as soon as possible. I am so sure you will comply with his Majesty's wishes, that I give myself the pleasure to wish you a good voyage.

The first problem against such urgency was a foreseeable delay for Boulton to read Poli's letter, which happened only at the end of September, also favored by the fact that Poli missed the appropriate recipient address (Chippindhall's 119 Fleet Street in London, instead of Truro in Cornwall). Nevertheless, Boulton promptly informed Watt of the novel, distinguished request from Naples on September 26,[10] and just two days after he answered to Poli in order to let him to realize the difficulty in designing and building an appropriate steam engine for the use required and, at the same time, to ask to reconsider the proportions of the engine requested:

> His Sicilian Majesty wishes to hire a New Engine of our invention erected near Naples of such power as to raise 10000 Cubic feet of water one foot high per minute; or say equal to the raising of 370 Cubic feet of water 27 feet high per minute which is more than three Millions of Cubic feet to work (for the Engine will work day & night if required). This is a large quantity of water, to water meadows. I therefore want you to reconsider whether you really have use for so much. [...]

---

[10] "We would furnish him with an Engine to raise 400 Cubic feet of water 27 feet high for his Majesty and pay us for the Engine but will make us a handsome reward & the offering to grant us a patent, also the King hath sent this month 2 Frigates to London & Mr. Poli desires I would return in one of them & bring the Fire Engine with me, & a workman to erect it". Matthew Boulton to James Watt, 26 September 1786. The Library of Birmingham: Birmingham Archives and Heritage. Boulton & Watt Collection: MS 3147/3/10.

These things I mention that you may see the real causes why I cannot so suddenly obey your summons & go aboard the Frigate you mention. [...]

I will consult my partner Mr. Watt so soon as I return to Birmingham which will be about the end of October [...]. In the intron, please to let me know the extent of the Meadow you mentioned (in English measure) & some particulars relative to the manner of watering it & how many hours per day you want have of Engine to work.[11]

Boulton's request for further details about the expected working of the steam engine pump for Carditello was certainly dictated by the impossibility for him to make a preliminary on-site survey, but was likely prompted as well by his and Watt's desire to accomplish in the best possible way the expectations of a client as illustrious as the Sovereign of one of the most important kingdoms in Europe. The effect of such request on the passionate researcher Poli was that, in order not to give approximate and hypothetical answers, he performed an accurate experiment to determine the quantity of water required for the irrigation of Carditello's fields, thus providing Boulton and Watt with a detailed explanation about the power of the steam engine to employ. On February 27, 1787 Poli wrote:

> The extent of the meadows is four hundred Moggios. One Moggio is $39990^{6/7}$ English feet squares, and it contains 46656 Neapolitan Palms square, and one and 1/6 Neapolitan Palm measures one English foot.
>
> The said 400 Moggios are to be divided in eight spots, each of fifty Moggios. One of these spots (that is 50 Moggios) is to be watered each day; so that all the 400 Moggios above mentioned shall be watered in eight days; and, of course, each spot of fifty Moggios shall be watered every fortnight.
>
> The Engine is to work fourteen hours a day, as the watering cannot be done but from six in the evening to eight in the morning, the heat of the day not permitting its being continued any longer without hurting the product of the earth.
>
> To give you an idea of the quantity of water, which is required for the said meadows I procured to get a standard in this manner. I have made my observations upon a little spot of land of the extent of five Moggios, which are watered once every fortnight; and having applied a large tube of about ten English Inches in diameter to the side of the Canal which leads the little stream employed to water the said five Moggios, I have occasioned all the stream to fall, through the said tube, into a cubical Receiver of 64 English cubic Feet, which was filled up in about 35 seconds of time. Now you must know that such a stream flows constantly during fourteen hours in order to water the five Moggios abovementioned: consequently, it will be required ten times as much water for the said fifty Moggios every day; or to say better eight time as much, allowance being made for the difference of the soil, which is not so bibulous, or spongy in the said 400 Moggios. Then you see Sir, what the Engine as mentioned, capable to raise 370 cubic feet of water per minute 27 feet high, is not sufficient for our purpose. However, His Majesty having been informed by me of all these particulars, told me that such an Engine will do very well; and that he does not want a larger one, as He intend to water so many Moggios only as showed want the quantity of water, which shall be raised by that Engine.
>
> The water raised by the Engine will run about one mile through a canal made of masonry; and then it will be distributed to the several spots of land through other canals disposed in the plain ground, so that the water overflowing from them will water the meadows as above.[12]

---

[11] Matthew Boulton to Joseph Poli, 26 September 1786. The Library of Birmingham: Birmingham Archives and Heritage. Boulton & Watt Collection: MS 3147/3/518/N02.
[12] Joseph Poli to Matthew Boulton, 27 February 1787. The Library of Birmingham: Birmingham Archives and Heritage. Boulton & Watt Collection: MS 3147/3/518/N03.

Being a court man, it is not surprising that Poli kept King Ferdinand IV constantly updated about the affair (also knowing about King's predilection for the *Reggia di Carditello*), and it would be as well not surprising, for the beloved (science) tutor of the young Crown Prince, to involve such illustrious pupil in the determination mentioned, as a practice lesson for him. A side-effect of this situation, however, was the urgency to have the engine, again expressed by the King, which forced Poli to reiterate to Boulton the request for a prompt delivery of it, along with the novel one for a skilled erector able to let the engine to work effectively and efficiently in a Country where no trained technicians were available:

> Our Frigates will sail for London next month in order to carry a present to His Britannic Majesty, and will stay there till the month of September. Therefore, His Majesty would wish that you was so good as to get ready the said Engine for that time, and put it on board the said Frigates. If you cannot come yourself with the Engine, you <u>must absolutely</u> send a person, who could set it at work, and instruct our people how to manage it at large; otherwise, it would be of no use, nobody here being acquainted with it.

### *4.3. A difficult start: Ramsden enters*

Notwithstanding the urgency expressed by Poli, several months passed without Boulton giving any answer, despite a novel letter forwarded to him in May 1787 through the agency of the Neapolitan Ambassador in London. The new summer season then arrived, and no news came from England about the engine for the King of Naples. On July 24 Poli wrote again to Boulton for a prompt response to his request, "His Majesty being very anxious to get such an Engine with the utmost speediness", by giving up a "professional" behavior and expressing a heartfelt appeal: "For God's sake, be so good as <u>not to fail to forward it</u>, as I promised to His Majesty that his order should be accomplished before the end of that year".[13] True to his Neapolitan style to leave no stone unturned, however, at the same time Poli sent also a letter to his old friend Jesse Ramsden in London (McConnell 2007), begging him to urge Boulton to answer. Ramsden, for its part, did not hesitate for a moment to take Poli's request into consideration, and already on August 16 promptly wrote to Boulton in Soho in a very convincing way:

> I know very well the many engagements you have and how every moment of your time is employed in objects of the greatest importance, otherwise I should feel myself very much hurt as my own disappointment as you promised to call on me when you was in London, but I know you man of the world think is not necessary to keep any engagements with an old woman, but I would not have you be too sure that you'll get off so easily for I assure you there is only one way I will forgive you, which is to answer this letter as soon as possible.[14]

This was not at all the only action undertaken by Ramsden, who managed to meet directly Boulton when he came on September 20 to London, where Boulton showed to Ramsden a sketch of a letter answering Poli, declaring, however, that he could not send an engine to Naples before the subsequent Spring. Of course, this was – anyway – a reassuring sign for Ramsden (and Poli), and Boulton was induced to promise to Ramsden to finish and send that letter to him on the next Saturday, in order for Ramsden to forward it to Poli. Probably unexpectedly, on that Saturday Boulton did not come to Ramsden's workshop at the Hay Market with any letter, but just saying he had been to the Neapolitan Ambassador in London without finding him in town, so that he could not have done anything on the business. At this point, knowing the need for Boulton to leave London for Cornwall the subsequent morning, he offered to deliver by himself any message for the Ambassador, but Boulton declined the offer with the further promise "upon his honour" to write a letter to Poli and send it to Ramsden before he left London, "though he might be obliged to set up all night, being very engaged that day". Needless to say, Ramsden did not receive any letter from Boulton, and no letter or message was collected by a servant of him sent two days later to Boulton's residence. The last, dramatic action put in place by Ramsden was then to write on October 1 an obliging letter to Boulton, by "threatening" him to send a burning letter to Poli (enclosed to his own letter to Boulton) where all the efforts performed by him to induce Boulton to answer were described in detail.

---

[13] Joseph Poli to Matthew Boulton, 24 July 1787. The Library of Birmingham: Birmingham Archives and Heritage. Boulton & Watt Collection: MS 3147/3/518/N05.
[14] Jesse Ramsden to Matthew Boulton, 16 August 1787. The Library of Birmingham: Birmingham Archives and Heritage. Boulton & Watt Collection: MS 3147/3/518/N06.

> I cannot forfeit my character nor the good opinion of my friends at Naples how as trifling you may think my application, be assured I will never give up my request till I can satisfy my friends that I have not neglected their requests or the commissions that they have entrusted me with.

> I therefore take the liberty to enclose a copy of a letter I intend to send to Mr. Poli [...]. I will wait fourteen days for your answer and if I do not receive any by that time I shall conclude you approve of it and I will send my letter to Naples by the neat mail after.[15]

The 14-days ultimatum was, this time, not disregarded and, one week later, from Truro in Cornwall, Boulton at last answered to Poli, by apologizing for the long delay (partly due to his absence on business in France and, anyway, for his commitment to business), and reassuring the customer about his interest in the engine for the King of Naples: "I must therefore entreat you to endeavour to appease his Majesty & to assure him that it is my greatest ambition to merit his approbation & that I will not fail to fulfil his Majesty commands early in the approaching year". The best guarantee for his genuine interest was that, in the same letter, Boulton finally came fully into the matter, including the price of the engine:

> I observe that you have made some experiments to determine the quantity of water wanted & that the result of those experiments is you want to have 370 English Cubic feet of water raised 27 English feet high. [...]

> I am persuaded it would be more agreeable to the King to pay one sum for all I will tell you as near as I can what I think it would come to fix: The Castings of Iron of the Cylinder [...] £ 1500; for Boulton & Watt drawings, directions trouble & various expenses & for their profit, altogether without any annuity as in England £ 1000, to which you must add the Wages of one Man from the time he sets out to the time he returns at about 25 Shilling per week & you must also add the freight to the place of its erection.[16]

At the same time, a memorandum was written by Boulton to Watt, with the determination of the technical details for the "New Engine":[17] the business definitively started.

## 5. *Poli's Engine* for the King of Naples

Come to the point, in that same letter Boulton asked Poli further details about the place where the engine was to be erected, in order to proceed in the best possible way for the preparation of all the necessary and appropriate parts of the engine. Poli provided the required technical details in a letter dated November 26 (written in an understandable detached way), where he enclosed also two plates containing the section and the plan of the river Volturno with the Royal Meadows at Carditello.

> Now having before you both the Section of the River, and the general Plan of the whole, you may see how to settle the matter touching the water course to be open from the River to the bottom of the Pumps; how it must be constructed, and to what a distance from the River it should be extended.[18]

---

[15] Jesse Ramsden to Matthew Boulton, 1 October 1787. The Library of Birmingham: Birmingham Archives and Heritage. Boulton & Watt Collection: MS 3147/3/518/N07.
[16] Matthew Boulton to Joseph Poli, 8 October 1787. The Library of Birmingham: Birmingham Archives and Heritage. Boulton & Watt Collection: MS 3782/12/6.
[17] [Matthew Boulton] to James Watt, [*sine data*]. The Library of Birmingham: Birmingham Archives and Heritage. Boulton & Watt Collection: MS 3147/3/518/N04.
[18] Joseph Poli to Matthew Boulton, 26 November 1787. The Library of Birmingham: Birmingham Archives and Heritage. Boulton & Watt Collection: MS 3147/3/518/N09.

**Fig. 3.** "List of the articles belonging to Poli's Engine for the King of Naples", that is, Boulton & Watt invoice for Poli's Engine (first page of four). Courtesy of *The Library of Birmingham: Birmingham Archives and Heritage. Boulton & Watt Collection*.

Poli also provided a number of suggestions about where the engine could be erected, how the water of the river conveyed, where the canal built and led, and so on. He got fully involved in the business, not only due to Boulton's request, but also for his own interest, which was only barely concealed behind the interest of the King "to be perfectly acquainted".

> You may set down your scheme at large upon the first Plate itself, and send it back again along with the Engine, or any other way you shall like the best.

> His Majesty wishing to be perfectly acquainted with the construction of the Engine in order to be able to judge by himself of anything relating to it, I beg you will send a full explanation of all its parts with the proper drawings annexed to it.

Probably, Poli's interest in the engine was twofold: on one hand the curiosity of the scholar might have prevailed, but on the other hand the teacher's diligence might have manifested once again, especially towards his young pupil, the Crown Prince Francis.

### 5.1. Setting the business

A not at all secondary issue, addressed in the same letter, was the economic one. Boulton asked a total of 2500 pounds for the engine to be built and for the profit for his and Watt's enterprise, but Poli pointed out that Boulton in person declared to him, when the two met in Birmingham years earlier, a total price of just 2000 pounds. Once again, the typical Neapolitan style came out:

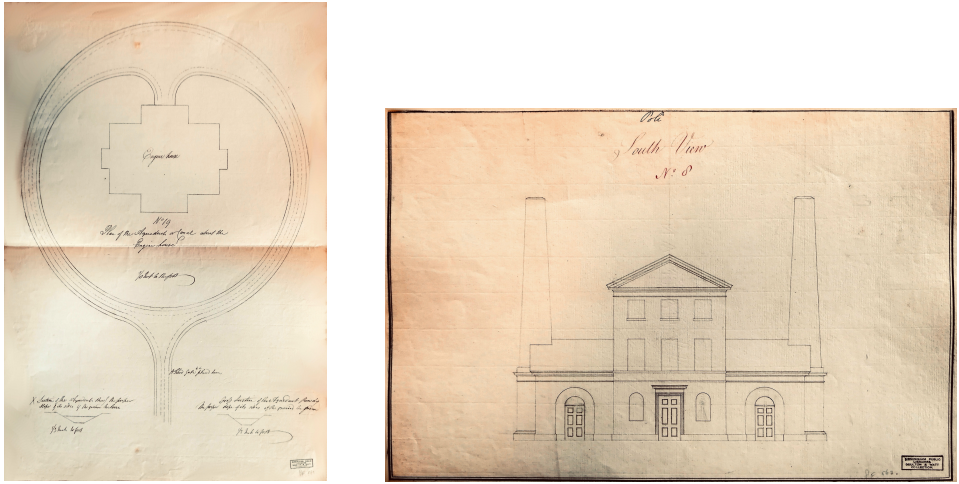

**Fig. 4**. House for Poli's Engine in Carditello. Left: "Plan of the Aqueduct or Canal about the Engine house". Right: South view of the Engine house. Courtesy of *The Library of Birmingham: Birmingham Archives and Heritage. Boulton & Watt Collection*.

> The only thing I would have you to consider is, that since the infancy of my project to His Majesty I gave Him a paper written of your own hand at Birmingham, when you was so good as to spend an evening there with me, and the Venetian Ambassador. There you set down all the dimensions, and prices of the Engine at large, with all the least particulars relating to it, so that it takes up a whole sheet of paper. The King therefore being already acquainted with those particulars in the most solemn way, I beg you will consider with your knowledge, and prudence, if it is honourable for me to mention any alteration about the price of the Engine. [...]
>
> His Majesty will give you one thousand Pounds for its price when delivered on board; besides one thousand Pounds for your profit, drawings, directions, trouble, and various expenses; without any annuity, and any other advantages of any kind. [...]
>
> Proper orders shall be given next week to the Neapolitan Ambassador in London, and you may apply to him for anything regards this subject.

In the first days of January 1788, the Neapolitan Ambassador in London, Count Lucchese, transmitted Poli's letter to Boulton, including the two plates mentioned above, and was officially appointed to take care of the business. Boulton then managed to see the Ambassador (finally succeeding in this only on the evening of February 4), with whom tried to settle the economic affair true to his English style:

> I know not what paper of particulars Mr. Poli alludes to which he says I gave him at Birmingham. I shall explain to the ambassador our usual terms & shall agree to show him the Invoice before the Engine is sent off which he may then accept or reject. Moreover the Engine we then talked of was a 33 Inch Cylinder to the best of my memory. Kings and ministers are so used to be imposed upon that they suspect all men.[19]

The issue was likely superseded (see below) by "all the dispatch possible" desired by the Ambassador, so that Boulton was left nothing to do but start with the production of the engine, and Watt was directly involved (of course) into it. In a couple of months all the pieces required were ordered and produced, and by the end of May 1788 an extremely detailed invoice (Fig. 3) was provided to the Ambassador by the Boulton & Watt company.[20] The price to be paid for only "cast iron, hammered iron and brass materials"

---

[19] Matthew Boulton to James Watt, 1 February 1788. The Library of Birmingham: Birmingham Archives and Heritage. Boulton & Watt Collection: MS 3147/3/12.
[20] Boulton & Watt Invoice, [*sine data*]: "List of the articles belonging to Poli's Engine for the King of Naples – Invoice of Cast Iron, Hammered Iron, and Brass materials for a Steam Engine furnished by Boulton & Watt of Birmingham in April and May 1788, to the

for the steam engine was at last about 2000 pounds: what Boulton envisaged earlier was evidently correct but, as we will see, the King of Naples (and Poli) did not displease at all the purchase done. Boulton "settled the plan" of the payment[21] with the Ambassador on May 29 and, one week later, the agent and banker of the Boulton & Watt company – Mr. William Matthews – "draw the Bill for the Ambassador to sign and it will accordingly be done next week, and Matthews will send it to Laghorn for payment".[22] Just in time before the engine embarked for sailing to Naples.

The business, however, was still not completely settled, and a number of troubles and concerns made their appearances. The most foreseeable ones regarded – of course – the carriage of the engine parts. Just to give a flavor of them, we report that yet on May 28, Boulton wrote (from London) to Watt that "Renney says the piston rod for Poli's Engine is ready, but there is no cap to it, as he says non-was ordered",[23] while the day after: "I believe the things from Hull are arrived and beg Mr. Roberts may be ordered to send everything from Soho directly by the Wagon. If there is any not quite ready, let them be sent by flying Wagon afterwards". On June 3, "neither the Goods from Chester nor Hall are yet arrived but I hope expect they will be in good time for the Ship that sails for Naples",[24] and on June 7 "neither of the Vessels are yet arrived from Gainsborough". The two Neapolitan frigates (likely, *Minerva* and *Cerere*) to be employed for the carriage of the engine, notably, had also another important task to complete (among the others), i.e., transporting purebred horses purchased by the lover King Ferdinand IV,[25] and this evidently explains the increase in concerns for Boulton & Watt.

A more interesting issue (for us) was, instead, that of the "drawings" of the engine, to which Poli was particularly interested in, as mentioned above, hiding himself behind the presumed King's desire "to be perfectly acquainted" about the steam engine (see Fig.s 4, 5, 6). Indeed, it is somewhat intriguing that both Boulton and the Neapolitan Ambassador, Count Lucchese, were unexpectedly anxious to deliver them to Naples with the engine. Of course, during the production and preparation of the different parts of the engine, the drawings prepared by Watt were owned by him, for whom it was quite usual to make changes and improvements until the very end, this being particularly true for the important order of the King of Naples (see above). Boulton was of course aware of this habit, so that it is not surprising that on May 29 he wrote to his partner: "I and Mr. Matthews are engaged to dine with the Ambassador on Friday 6th June viz tomorrow sevenight, and therefore you may keep the drawings a few days longer." Two days later, however, on May 31, kindness already gave way to anxiety: "Please to remember that I am engaged to dine with the Ambassador next Friday and Matthews also, where I will settle all money matters & deliver the drawings".[26] Three more days later: "I hope the drawing will arrive time enough for me to keep my promise of presenting them to him on Friday when I am to dine with him".[27] Fortunately for Boulton, in the end everything went well, and he delivered all the drawings to the Ambassador in due time, keeping Watt duly informed of the outcome of the June 6 dinner.[28]

---

order of Joseph Poli Esq. on account and risque of". The Library of Birmingham: Birmingham Archives and Heritage. Boulton & Watt Collection: MS 3147/3/518/N10.

[21] Matthew Boulton to James Watt, 29 May 1788. The Library of Birmingham: Birmingham Archives and Heritage. Boulton & Watt Collection: MS 3147/3/12.

[22] Matthew Boulton to James Watt, 7 June 1788. The Library of Birmingham: Birmingham Archives and Heritage. Boulton & Watt Collection: MS 3147/3/12.

[23] Matthew Boulton to James Watt, 28 May 1788. The Library of Birmingham: Birmingham Archives and Heritage. Boulton & Watt Collection: MS 3147/3/12.

[24] Matthew Boulton to James Watt, 3 June 1788. The Library of Birmingham: Birmingham Archives and Heritage. Boulton & Watt Collection: MS 3147/3/12.

[25] Matthew Boulton to James Watt, 29 May 1788, *cit*.

[26] Matthew Boulton to James Watt, 31 May 1788. The Library of Birmingham: Birmingham Archives and Heritage. Boulton & Watt Collection: MS 3147/3/12.

[27] Matthew Boulton to James Watt, 3 June 1788, *cit*.

[28] Matthew Boulton to James Watt, 7 June 1788, *cit*.

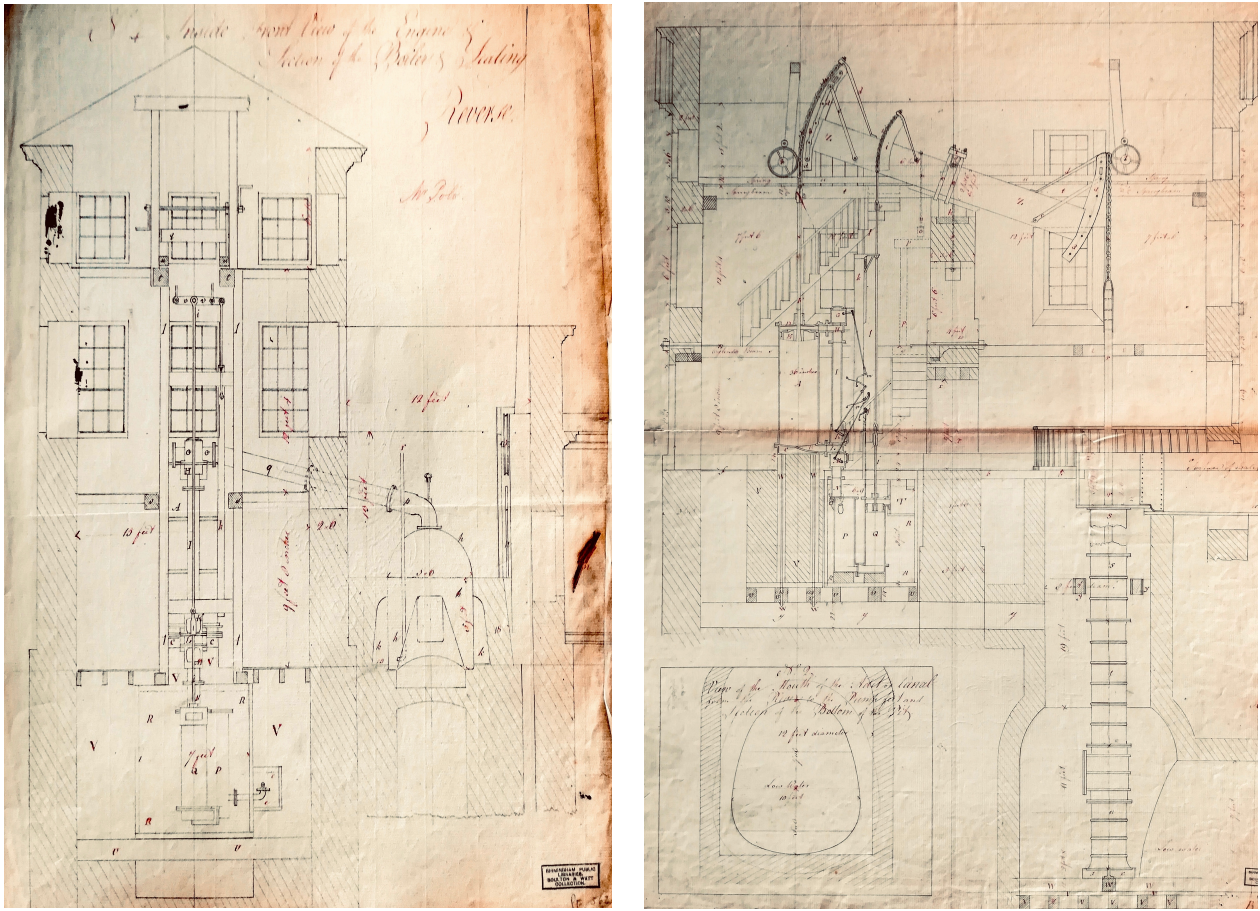

**Fig. 5.** Poli's Engine in Carditello. Left: "Inside Front View of the Engine & Section of the Boiler's seating". Right: "N.1: General section"; "N.2: View of the mouth of the Aqueduct or Canal from the River to the Pump foot and Section of the Bottom of the Pit". Courtesy of *The Library of Birmingham: Birmingham Archives and Heritage. Boulton & Watt Collection.*

### *5.2. A double-acting engine*

As mentioned above, *Poli's engine* was designed explicitly by Watt, although it shared some standard features with other engines Boulton & Watt built in the same period. The engine, indeed, was based on a double-acting cylinder, where the steam acted alternatively on both sides of the piston. It also included the *parallel motion* mechanism devised by Watt for such engines, where a mechanical linkage with four bars coupled with a pantograph allowed the (approximately) rectilinear up-and-down motion of the piston rod to be transmitted to a beam moving in a required circular arc. Nevertheless, Watt introduced some exceptional features as well in the engine he prepared for the King of Naples, when compared with other engines (Dickinson & Jenkins 1927).

First of all, the working gear had an arbor for each valve (steam, equilibrium and exhaust valves), with the gears of these valves being interlocked (see Fig. 6). More specifically, the opening of the equilibrium valve was subject to the closing of the exhaust valve, while the opening of the steam and exhaust valves was subject to the closing of the equilibrium valve. The injection valve was, instead, independently worked. The steam and equilibrium arbors each had a detent adapted to engage a pivoted catch, each tending to turn in the direction to open its own valve by means of a weight suspended upon an arm projecting from it. Some leather straps connected the tail of the top arbor detent catch to a lever on the equilibrium arbor and the tail of the equilibrium arbor detent catch to a lever on the exhaust arbor, while another lever on the exhaust arbor to a lever on the top arbor.

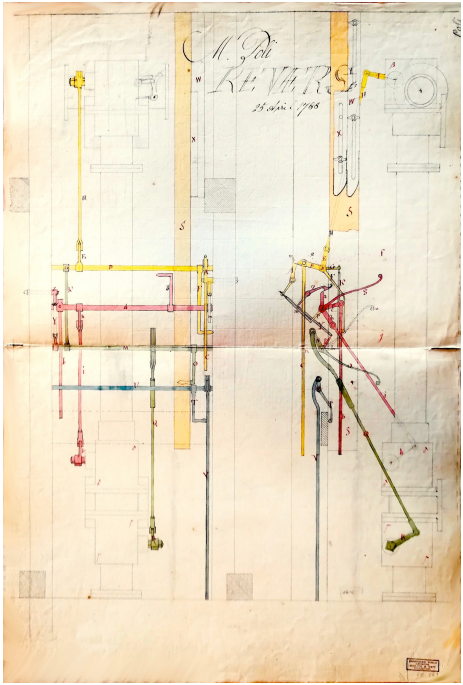

**Fig. 6.** Working gear for Poli's Engine, with the steam arbor (red), equilibrium arbor (green) and exhaust arbor (blue). Courtesy of *The Library of Birmingham: Birmingham Archives and Heritage. Boulton & Watt Collection*.

When the equilibrium arbor rotated to close its valve, thus shutting off the communication between the top and the bottom of the cylinder, the catch of the top arbor detent was released, the top arbor turning by means of the suspended weight to open its own valve and pulling round (through a strap) the exhaust arbor to initiate the movement of its valve. Such movement was completed by the steam pressure on its upper face, while the valve carried with it the exhaust arbor and the attached parts: in such a way, the top of the cylinder was in communication with the boiler and the bottom with the condenser, the downstroke of the piston ensuing. In the corresponding downward movement of the plug-tree, a chock acted upon the handling arm of the top valve, while another chock acted upon that of the exhaust valve (which were now in their raised positions), thus restoring them in the position shown in the Figure with the top and exhaust valves shut. Instead, when the exhaust arbor rotated by means of a lever and a strap, pulled down the detent catch of the equilibrium arbor, this turning by means of the suspended weight to throw open the equilibrium valve for the up-stroke. As the piston ascended, a peg on the plug-tree engaged the lower side of a handling arm, raising it to close the equilibrium valve: with the arm restored to the position shown, it pulled a strap to disengage the catch of the top arbor detent for the downstroke.

Injection in the engine was worked by an arm and a rod: the valve was opened by means of a peg on the plug-tree on the underside of the arm, while closed – as the piston began to descend – by chock pressing upon its upper side. The equilibrium valve was opened when the corresponding arbor was released by the rotation of the exhaust arbor; the top gear was released to open the valve by the rotation of the equilibrium arbor, and the exhaust arbor was turned to start the opening of the exhaust valve by the rotation of the top arbor. Finally, each gear was restored to the position shown in the Figure by the action of the plug-tree.

### 6. Serving His Sicilian Majesty

It was customary for Boulton & Watt to send a trained technician to erect the engine when it became no longer possible for them to move from their headquarters. This issue was addressed as early as in January 1788, when the Poli's engine business was definitively settled, and Boulton proposed to Watt to send one of their best collaborators to Naples: "I think [Malcom Logan] is the prepared man we have to send to

Naples as he is a handy fellow either in Wood or Iron or Engine or Mill or pumps".[29] Malcom Logan was part of the squad of Scotsmen working for Boulton & Watt since 1781, a mechanic of quick intelligence to whom Boulton refers as "a handy, active and industrious fellow" (Dickinson & Jenkins 1927).

## 6.1. *A prepared engine man*

Unfortunately, the industrious Logan soon gave himself to drunkenness, as it was practically customary for the average English workman of that day: when sent to a distance to fit up engines, mechanics were left in a great measure to themselves, so that they were apt to become careless and ill-conditioned. Notwithstanding this, in 1785 Boulton & Watt entrusted Logan with the important task to erect a steam engine (1785-1787) for land drainage for the Batavian Society in the Netherlands, although Logan's "disease" manifested itself also in this occasion. On February 24, 1786, indeed, Watt expressed all his regret to Boulton, who appears as Logan's true sponsor: "I am very sorry to hear that Malcom Logan's disease increases. I think you should talk to him roundly upon it, and endeavour to procure him to make a solemn resolution or oath against drinking for some given term" (Smiles 1865). Nevertheless, Logan turned out to be a very skilled mechanic, even able to make improvements to the original Watt's design and demonstrating the power and flexibility of the engine erected on the Blijdorp polder in the province of Holland – for instance – by shutting down the injection water for the condenser and letting the engine run for a couple of minutes "on the air pump" (Roberts 2004; Verbruggen 2005; Dickinson & Jenkins 1927).

Logan's experience in Holland was very troubled for different reasons. Before the steam engine finally proved to be capable of flushing the Rotterdam canals, Watt wrote to the secretary of the Batavian Society – a savant-fabricant – Jan Daniel Huichelbos van Liender in April 1787 that he "wanted to have dispatched Malcom Logan but he told me that he was not yet quite able to go on account of a very dangerous fistula in and which he had almost died of" (Verbruggen 2005). Watt was, indeed, "extremely vexed that M. Logan has not been able to set out sooner, as it deranges us in our business here, having more to do than we have men to execute & having few who are so fit for your business or any other which requires judgement as he is, though he be a drunken fellow he is honest & ingenious." Later, in October of the same year, when the Dutch engine proved successful, van Liender wrote to Watt that he was "very well contented with his abilities, but I am sorry for himself, that he is so little master of his passions for that low vice of inebriety. He is quite addicted to it, and it will always prove his ruin; a great pity it is, that a man with such fine parts, is so much a slave of that misconduct" (Verbruggen 2005). The same van Liender, however, proved to be not at all "sorry for himself" when, four years later, an experienced man was required for drainage at Meydrecht: he did come back to Watt, writing that "if Logan was returned from Naples, as he has been already once in Holland, I should think best to send him over, and if he is somewhat cured of his drinking fault, it would be so much better" (Verbruggen 2005).

All these facts, if they induced Boulton to suggest employing the "honest & ingenious" Logan to erect Poli's engine, likely prevented Watt to adhere enthusiastically to the proposal of sending a "drunken fellow" to serve the King of Naples. In May 1788, indeed, Watt proposed to employ James Lawson rather than Logan, and this alternative offer was also accepted by the Neapolitan Ambassador, promptly informed by Boulton:

> The Ambassador agrees to take Lawson upon the same terms as Logan via 25 per week & pay his passage. I think is possible that Lawson may promote the Erection of Engines in the other Italian States but I think we should agree with him upon terms that may promote our interest as well as his non-otherwise not send him. I therefore wish you would talk with him before you consent to his going & if you agree I then think it right that you should give him a Lesson upon Canal making as well as upon the minutia of the Erecting of the Engine.[30]

---

[29] Matthew Boulton to James Watt, 25 January 1788. The Library of Birmingham: Birmingham Archives and Heritage. Boulton & Watt Collection: MS 3147/3/12.

As well as Logan, Lawson was one of the staff of engine erectors in Cornwall since 1782, and Boulton's judgment[31] that he could "manage the working of an engine better than any of their engine men" (Dickinson & Jenkins 1927) certainly convinced the Count Lucchese to recruit Lawson instead of Logan. However, the experienced Ambassador probably suspected some possible "embarrassment" in Naples due to this change of people and, just two days later, suddenly changed his mind in order not to be in a bad position towards the King of Naples: "I have just received a message from the Neapolitan Ambassador saying that upon looking over his Letters etc. he finds it would be improper for him to send Lawson".[32] Informed by his partner, Watt was not pleased by such intrusion, but the diplomatic Boulton readily realized that insisting on this point would have been inappropriate for them and "if we do it will be upon our risk and must rest upon the honour of the King, which is very precarious".[33] The end of this story was that on June 24 a letter was prepared by the Boulton & Watt Company to be presented to Poli by the technician Malcom Logan, when arrived in Naples:

> In consequence of your obliging correspondence with our M. Boulton, we have prepared & delivered to His Excellency Count Lucchese's order in London, a steam Engine Cylinder of 36 inches diameter, a suitable pump, a boiler & condenser & all the other cast Iron, hammered Iron, Brass & Copper apparatus necessary to complete the Engine. We have also delivered to His Excellency a complete set of drawings & directions for the erection, which we hope will sufficiently explain the structure of the House.
>
> We have sent all the expedition in our power, yet it has been with difficulty that we have been able to complete & fetch the materials here by this time. We have made every part complete as far as could be done here & are certain there is no article of importance which have been omitted to be sent, though perhaps some small articles may have been neglected, as it is impossible to recollect or to have lists of all the minutia of such an Engine. We send this by Malcolm Logan an experienced workman who have erected many Engines is ingenious & intelligent & are expected well do his business to His Majesty & your satisfaction. It has been with great difficulty we have prevailed with him to go so far from home but doubt not of his exerting himself to complete the business quickly & well. He has been employed in collecting & helping to ship the Goods from the beginning of last weeks, at which period we agreed with him that his services to His Sicilian Majesty should commence.[34]

### 6.2. Striking success in Naples

Notwithstanding Logan's skills, the erection of Poli's engine for the Royal Estate of Carditello required some a long time (more than one year) for different (though usual) technical problems but, at last, both the functioning of the engine and Logan's work was generally very well appreciated, and the satisfaction was complete. The credit company (operating in Naples) of the merchants Charles Cutler and Christian Heigelin, who were correspondents from the Bourbon capital of numerous English citizens, on September 1790 duly informed Boulton & Watt that Logan gave "great satisfaction in his situation here, and of which His Majesty has given frequent testimonies" and, even more relevant for the two businessmen, that "he also wishes to have your answer about the Steam Engines, as he says 1 or 2 more will be wanted".[35]

---

[30] Matthew Boulton to James Watt, 29 May 1788, *cit*.
[31] About the choice of the man to be sent to Naples, the businessman Boulton was probably more concerned to promote the interests of the Company than to select a skillful technician, as evident from the letter quoted.
[32] Matthew Boulton to James Watt, 31 May 1788, *cit*.
[33] Matthew Boulton to James Watt, 3 June 1788, *cit*.
[34] Boulton & Watt Company to Joseph Poli, 24 June 1788. The Library of Birmingham: Birmingham Archives and Heritage. Boulton & Watt Collection: MS 3147/3/518/N11.
[35] Cutler & Heigelin to Boulton & Watt, 11 September 1790. The Library of Birmingham: Birmingham Archives and Heritage. Boulton & Watt Collection: MS 3782/12/35/162.

We do not know who requested novel engines to Boulton & Watt but, as a matter of fact, after the completion of the erection of the steam engine working for supplying water to the Royal Estate of Carditello, Logan was asked by the general manager of the Bourbons – Saverio Guarini – for that estate to prepare hydraulic engines for the *Molini di S. Antonio* (St. Anthony Mills; see Fig. 2) on the *Lagno maggiore* river. This mill was to serve for grinding the cereals grown in the Carditello's estate using eight wheels, two located on the ground floor and six on the upper floor, driven by hydraulic engines (Serraglio 2003). The building of the new mill was approved by King Ferdinand IV on July 17, 1792 but, some time before, another steam engine was requested to Boulton & Watt to be erected in the Spain ruled by the same Bourbon family. Indeed, in June 1792 a blowing engine finally arrived in Cadiz, as agreed by Boulton & Watt with the Spanish naval officer Fernando Casado de Torres who, as early as June 1788, suggested to the Spanish Navy a proposal for a fire engine intended to work a sawmill in the *Arsenal de La Carraca* (Quijada 1998). That proposal was, probably, independent of Poli's request to Boulton & Watt, but it is quite remarkable the coincidence that the final contract was stipulated just in 1790, when Poli's engine finally revealed its success; the Cadiz engine arrived in Spain only in 1792.

Of course, also in Cadiz an erector was required, and, for some reason, Boulton & Watt again chose Malcom Logan for this aim, convinced that his presence in Naples was now not at all essential. Boulton, however, probably foresaw some possible obstacle due to the personal success of Logan in Naples (maybe informed by Logan himself), and on July 4, 1792 wrote a letter to Poli requesting his own intervention in order to let Logan to leave from Naples, by using any kind of persuasion at his disposal:

> I want request the favour of your good offices with his Majesty Ministers or any be proper to allow Malcom Logan to leave Naples & go by the first Ship to Cadiz in Spain where we have delivered a Steam Engine for the use of his Catholic Majesty.
>
> When Malcom hath erected this Engine, he may return again to Naples in case he is wanted.
>
> We fear we shall disappoint his Catholic Majesty unless Malcom is allowed to depart & go by the first Ship to Cadiz which I beg you will accelerate by every means in your power.
>
> If I can render you any services in this Country I shall be happy to have it in my power.[36]

Also, to be on the safe side, the following day he wrote as well to the powerful English Ambassador in Naples, William Hamilton, asking his personal "interference" in the affair (but including the letter to Poli):

> We are unwilling to send another man there [...]. We wrote Malcom a fortnight ago, & now write to him again this day & enclose to him a letter of introduction & Credit to a Merchant in Cadiz.
>
> If any difficulty or obstruction to his departure should arise I must request the favour of your kind interference both in regard to the Neapolitan Government or any other that may impede his departure by the first Ship for Cadiz.[37]

On July 31, Hamilton promised to Boulton that "no pains of mine shall be spared to serve you", and, if he discouragingly pointed out that Logan "is looked upon here as a very great man and they increase his pay daily to induce him to remain here", nevertheless prudently informed Boulton as well that "I know he has instructed a Mr. White in what relates to the Engine which makes me hope that they may let him go".[38] Notwithstanding Hamilton's intervention, still many months passed without success. On February 21, 1793 James Watt Jr. – who was traveling to Italy – informed his father that Logan once again delayed his departure for Spain in order to finish his work for the Mill at Carditello.[39]

---

[36] Matthew Boulton to Joseph Poli, 4 July 1792. The Library of Birmingham: Birmingham Archives and Heritage. Boulton & Watt Collection: MS 3782/12/37/114.

[37] Matthew Boulton to William Hamilton, 5 July 1792. The Library of Birmingham: Birmingham Archives and Heritage. Boulton & Watt Collection: MS 3782/12/37/117.

[38] William Hamilton to Matthew Boulton, 31 July 1792. The Library of Birmingham: Birmingham Archives and Heritage. Boulton & Watt Collection: MS 3782/12/37/141.

Indeed, an actual problem arose for the effective working of the *Molini di S. Antonio*, since the driving force provided by the *Lagno maggiore* river had proved insufficient to simultaneously operate the six millstones on the top floor of the mill, forcing the operators to use them alternately. The expert Malcom Logan was, then, asked (along with Francesco Collecini and Carlo Pollio, among the others) to consider that problem, which was referred to a technical commission appointed to evaluate the possibility of introducing water from other neighboring rivers (Serraglio 2003).[40] Notwithstanding this, around March 1793 the Mill in Carditello was completed, and the same James Watt Jr. (who, in the meanwhile, met Logan) ingenuously wrote to his father about his optimism concerning Logan's leave for Spain. To this regard, it is quite interesting to know how the Neapolitan authorities proceeded in that occasion to delay (or even avoid) Logan's leave, as described by Logan himself:

> The King called for all my Letters, with an intention to write immediately to you, requiring your consent to my staying here, and for send me some other person in place. The Letter you addressed to the Intendant of the Marine at Cádiz, they also had from me and have never returned it me must therefore has the favour of you not to omit favouring me with another Letter for the Intendant […]. It is necessary Gentlemen I should acquaint you, that the Leave the King has given me for going on your Business to Cadiz is on Condition of my having given my Word & promise to return here as soon as your Business there be completed. Everything that depends on me, you may confidently rely on being executed with the greatest attention, and in such a manner as shall do Honour to their Employers, to you and myself.[41]

Finally, on January 1794 Logan departed for Spain, where he found a disastrous situation, to say the least. Indeed, the pieces of the steam engine arrived in 1792 showed some signs of deterioration because of the humidity and the poor conditions in which they had been stored, awaiting the construction of a new building to place the engine, since the previously predetermined location to house it proved to be not suitable due to a lack of solid foundations. Also, sometime after Logan's arrival, in 1795, there was the theft of an important quantity of pieces of the steam engine and the mechanism of saws, so that both were incomplete and unable to work. The technician and the Spanish officer Casado de Torres tried to replace the stolen pieces, but they had no room for it. As probably expected, however, these were not the only problems for Logan to keep at bay, given the renewed urgency of the Neapolitan Government who pawed to get him back in Naples, as recounted by Logan himself in a letter to Watt:

> I have been called on by the Neapolitan Consul by order of the Minister at Naples to know for what case I stayed so long. The following is the copy of a letter I sent to him as I was ordered to write to him: […] "I have been unluckily detained a great deal longer than I expected owing to a variety of unforeseen Accidents the principal of which is caused by a mistake committed long before I came in the foundation of a Building destined for the reception of an Engine I was to put use in the [Carraca] and which renders it absolutely necessary to begin the work anew before I can proceed any further in the Business, and my chief motive for wishing to stay and complete the same is the regard I have for my former Employers and Manufacturers of the Engine who would think their Reputation injured were it not to be properly finished, but as my greatest Desire is to return to Naples as soon as […] I have order to that effect and permission from the Court of Spain".[42]

The need to have the expert technician in Naples was, expectedly, genuine, as well as Logan's "greatest desire to return to Naples", but this last one was perhaps due to a reason not reported in the letter above, but rather into another one written by Casado de Torres to Boulton & Watt in order to inform the English Manufacturers of the situation in Cadiz:

> Lorsque je reçus votre lettre du 24 Décembre je passai office à M. l'Intendant de ce Département pour qu'il sollicita de la Cour de Naples de proroger le congé de Malcom; et je sais que la négociation est ouverte par le moyen de l'Ambassadeur, et même je crois qu'il n'y aura pas de difficulté pour l'obtenir.
>
> Malgré tout ce qui je viens de vous dire de présenta avant hier matin Malcom Logan en dissent qu'il nous laisserait sans achevés l'ouvrage, et qu'il s'en irait tout de suite si nous ne lui donnions les 14 guinées par mois qu'on lui donnait à Naples. E bien, Messieurs, a cet insulte grossier de Malcom nous avons répondu que nous lui donnerai les 14 guinées par mois, puis qu'il nous métrait le couteau à la gorge. Vous avons, donc, votre Malcom Logan avec 14 guinées par mois, avec sur logement, bois, eau, servant, et en fin avec tout ce qu'il a demandé: mais sera t'il content encore? Je crois que non.[43]

---

[39] Boulton & Watt to Fermin de Tastet & Co, 25 March 1793. The Library of Birmingham: Birmingham Archives and Heritage. Boulton & Watt Collection: MS 3147/5/592.
[40] This project was initially supported by the Neapolitan government but, later, was blocked, because it was considered harmful for the cultivation of hemp and flax.
[41] Malcom Logan to Boulton & Watt, 31 December 1793. The Library of Birmingham: Birmingham Archives and Heritage. Boulton & Watt Collection: MS 3147/3/395.
[42] Malcom Logan to James Watt, 28 January 1795. The Library of Birmingham: Birmingham Archives and Heritage. Boulton & Watt Collection: MS 3147/3/415.

For one reason or another, Logan did stay in Spain for less than two years, and eventually came back to Naples, where some more works for the mills in Carditello were waiting for him. The Bourbon Government did not miss the opportunity to have a good, qualified mechanic at his disposal to improve more and more the conditions of the Kingdom of Naples, along with its prestige.

**7. Conclusions**

The intriguing story, recounted above, concerning the introduction in Italy (through the Kingdom of Naples) of the most important technological innovation of the XVIII century, i.e., Watt's steam engine, has unfortunately a depressing ending, especially for the history of the Kingdom of Naples. Indeed, "the great fire engine of 3 feet in diameter erected on the Volturno river near Capua, which raised 30 thousand cubic feet of water per hour to 25 feet high, thus feeding a canal extending to Carditello to irrigate those campaigns" was destroyed "for the disorders of 1799",[44] when the Parthenopean Republic was established during the French Revolutionary Wars, after King Ferdinand IV fled before advancing French troops. James Watt was informed by his son Gregory – who in mid-1801 set off on European travels in hopes of recovering his health, then becoming a geologist and mineralogist – about what happened during those days, leaving for some hope to establish again the destroyed engine:

> The country people have carried off everything from the engine at Caserta except the cylinder and pumps and they even attempted to blow up the [engine] house. All this will need renovation and when the King returns in a few weeks it will be set about and Malcom Logan will send an order to you and demand a list of prices. Of course, if anything is done it will have unexceptionable security, for the Kingdom of Naples is ripening for Hell.[45]

Unfortunately, contrary to what claimed by Gregory Watt, the first steam engine in Italy was not restored, and the same acclaimed mechanic Malcom Logan ended tragically his days, dying in indifferent circumstances shortly after the French invasion of the Kingdom of Naples in 1806.[46] As a contradiction that is often repeated in history, the Republican rebels of 1799, who asked the Bourbons for an improvement in their living conditions, saw in that technological progress only a value to be cannibalized immediately. As a consequence, even the historical memory of this first attempt at technological modernization of the Kingdom of Naples was soon lost, and this explains the absence of such an important episode also in history textbooks. These constantly report, indeed, that the first application of the steam engine was realized in Italy only 30 years later, when the Neapolitan shipyards in Vigliena built the first steamship, honoring the same King who introduced the steam engine in Carditello: the *Ferdinando I* sailed from Naples for his maiden voyage to Genoa and Marseilles on September 27, 1818 (Ressmann 2007). It was not a joke of fate that, still in the Kingdom of Naples the steam propulsion had to be introduced, first in the whole Mediterranean. Indeed, as in a letter to Boulton & Watt by the Sardinian Minister, Marquis Grimaldi (who was the catalyst for such an introduction),

> there is no Country like the Two Sicilies that offers many advantages for the employment of Steam-boats: both in the Adriatic and Mediterranean the navigation can be performed coast-ways, as well as from Naples to Sicily and Palermo, the coasts not being rugged, and presenting various landing places. The Company likewise enjoys the exclusive privilege of providing whatever Steam-engines might be required for Manufactories and Mills, in the two Kingdoms; many of which would be necessary as several Provinces from want of water are obliged to send their oil to be pressed and their corn to be grown at Naples: when once the steam-engines were introduced many Manufactories would be established, and the steam-engines might at the same time be employed in the Mills.[47]

---

[43] Fernando Casado de Torres to Boulton & Watt, 3 June 1795. The Library of Birmingham: Birmingham Archives and Heritage. Boulton & Watt Collection: MS 3147/3/421.
[44] See the *Rapporto generale sulla situazione delle strade, sulle bonificazioni e sugli edifici pubblici dei Reali domini* (General report on the situation of roads, drainage and public buildings of the Royal domains) in the *Giornale del Regno delle Due Sicilie*, May 3, 1827.
[45] Gregory Watt to James Watt, 27 May 1802, reported in (Torrens 2006).
[46] James Watt Jr. to Arthur Oughterson, 10 January 1807, reported in (Tann 1978).

This is, however, another (yet intriguing) story, to be told elsewhere, which apparently did not benefit of the intervention of Giuseppe Saverio Poli. Nevertheless, we cannot help noting here that, although the architect of the realization of the first steam engine in Italy was *de facto* not directly involved in the realization of the first steamboat in the Mediterranean, it would be unwise to completely exclude any role of his even in such further technological improvement of the Kingdom of Naples. Indeed, still in 1817 (and even later) Poli's influence on King Ferdinand was certainly relevant, especially when concerning science and technology. This was only marginally due to his friendship with his former pupil, the Crown Prince Francis, while resting on his fame of a solid scientist and a reliable manager, as well as on his devotion to the Bourbons, whose major manifestation in that period was certainly his life chairmanship of the *Reale Istituto d'Incoraggiamento alle Scienze Naturali.*

Though not belonging to that densely inhabited network of entrepreneurial engineers which swarmed the second half of the eighteenth century, Poli nevertheless drove a certain productive circulation of both knowledge and innovation throughout Europe, with significant results obtained in Italy and, especially, in the Kingdom of Naples. His activity (and ability) in spreading scientific knowledge, as well as in fostering technological innovation derived from it, was certainly not at all unique in Europe (and, to a certain extent, he learned from his associates abroad), but it proved to be quite crucial for his Country and, more in general, for Italy. Indeed, Poli fully embodied the Enlightenment ideals of utility and progress and was able, by means of his strict relationships with the Bourbon court, to translate those ideals into practice, a shining example of this having been the one recounted in the present work. It is certainly true that, just as people throughout Europe continuously looked to England for inspiration, the Kingdom of Naples closely followed Great Britain in developing the modernization of the State. This happened, however, with the simultaneous "discovery" – able to increase the national prestige – by English (and others) of an immense and little known cultural and natural heritage (including Roman and Greek ruins, the marvelous Mount Vesuvius, Sicily, and so on) that largely called for the *Grand Tour* across the Kingdom of Naples. The lesson learned by historians simply recognizes the crucial importance of tending to regional specificities and differences, even when accounting for the history of industrialization. In this respect, Poli's Engine was just one example of that British industrialization that relied on global interconnections among people, products and processes (Misa 2004).

The beneficial effects produced by Poli's long lasting action in promoting Science in his Country (and even abroad, through his published works) endured well beyond his death and brought a deserved fame to the Kingdom of the Two Sicilies as the most technologically advanced state in Italy, and one of the most influential one in Europe. Unfortunately, the Bourbons were not so far-sighted to extend such benefits to any social class of their Kingdom, and the destruction of the first steam engine in Carditello was a painful example of that politics, which unluckily produced – in the present case – even the cancelling of the historical memory of that technological prodigy.

---

[47] Marquis de Grimaldi to Boulton & Watt, 25 March 1817. The Library of Birmingham: Birmingham Archives and Heritage. Boulton & Watt Collection: MS 3147/3/439.

## Acknowledgments

The present work was stimulated some years ago by Roberto Mantovani and Giuseppe Saverio Poli IV: my sincere thanks are here graciously expressed. I am also indebted to Andrea F. Scalella and Giovanna P. Perdonà for their invaluable help in archival searches. The very precious collaboration of the staff of the Birmingham Central Library (Birmingham), British Library (London) and Biblioteca Nazionale "Vittorio Emanuele II" (Naples) is kindly acknowledged as well.

## References

Andrew J. (2009). "The Soho Steam Engine Business", in Mason S., *Matthew Boulton: Selling What All the World Desires*. New Haven: Yale University Press.
Croce B. (1992). *Storia del Regno di Napoli*. Milano: Adelphi.
Della Porta G.B. (1606). *I tre libri de' spiritali*. Napoli: Iacomo Carlino.
Dickinson H.W. and Jenkins R. (1927). *James Watt and the steam engine: the memorial volume prepared for the Committee of the Watt centenary commemoration at Birmingham 1919*. Oxford: Clarendon Press.
Dickinson H.W. and Vowles H.P. (1948). *James Watt and the Industrial Revolution.* London: The British Council.
Esposito S. (2017). "Enlightenment in the Kingdom of Naples: the legacy of Giuseppe Saverio Poli through archive documents", in Esposito S. (ed.), *Proceedings of the 36th Annual Congress of the Italian Society of the Historians of Physics and Astronomy* (Naples, October 4-7, 2016). Pavia: Pavia University Press.
Esposito S. (2019). "Darwin and the others: the reception of Poli's *Testacea* outside Italy and other recent discoveries about the Molfetta scientist", in Garuccio A. (eds.), *Proceedings of the 37th Annual Congress of the Italian Society of the Historians of Physics and Astronomy* (Bari, September 26-29, 2017). Pavia: Pavia University Press.
Esposito S. (2020). "Giuseppe Saverio Poli e lo sviluppo della scienza tra la fine del Settecento e l'inizio dell'Ottocento". *Rendiconti dell'Accademia Nazionale delle Scienze detta dei XL - Memorie e Rendiconti di Chimica, Fisica, Matematica e Scienze Naturali*, 1, pp. 125-139.
Giustiniani L. (1797). *Dizionario Geografico Ragionato del Regno delle Due Sicilie.* Naples: Manfredi.
Hunter L.C. and Bryant L. (1991). *A History of Industrial Power in the United States, 1730–1930, Vol. 3: The Transmission of Power*. Cambridge, Massachusetts: MIT Press.
Landes D.S. (1969). *The Unbound Prometheus: Technological Change and Industrial Development in Western Europe from 1750 to the Present*. Cambridge: Press Syndicate of the University of Cambridge.
McConnell A. (2007). *Jesse Ramsden (1735–1800): London's leading scientific instrument maker.* Aldershot: Ashgate.
Misa T. (2004). *Leonardo to the Internet: Technology and culture from the Renaissance to the present.* Baltimore: Johns Hopkins University Press.
Piccari P. (2007). *Giovan Battista Della Porta. Il filosofo, il retore, lo scienziato*. Milano: Franco Angeli.
Poli G.S. (1781). *Elementi di Fisica Sperimentale*. Napoli: Fratelli Raimondi.
Poli G.S. (1791). *Testacea Utriusque Siciliae eorumque Historia et Anatome tabulis aeneis illustrata. Tomus Primus*. Parma: Ex Regio Typographeio.
Poli G.S. (1794). *Elementi di Fisica Sperimentale*. Venezia: Tipografia Pepoliana.
Poli G.S. (1795). *Testacea Utriusque Siciliae eorumque Historia et Anatome tabulis aeneis illustrata. Tomus Secundus*. Parma: Ex Regio Typographeio.
Poli G.S. (1826). *Testacea Utriusque Siciliae eorumque Historia et Anatome tabulis aeneis illustrata. Tomus Tertius. Pars Prima Posthuma*. Parma: Ex Ducali Typographeio.

Quijada J.H. (1998). *Transferencias de tecnologia britanica a comienzos de la revolucion industrial*, in Hourcade J.L.G. *et al.* (eds.), *Estudios de Historia de las Tecnicas, la Arqueologia industrial y las Ciencias*. Segovia: Junta de Castilla y Leon.

Ressmann C. (2007). "La prima nave a vapore del Mediterraneo". *Rivista Marittima*, February issue.

Rizzi Zannoni G.A. (1812). *Atlante Geografico del Regno di Napoli*. Napoli: [s.n.].

Roberts L. (2000). "Water, steam and change: the roles of land drainage, water supplies and garden fountains in the early development of the steam engine". *Endeavour,* 24, pp. 55-58.

Roberts L. (2004). "An Arcadian Apparatus: The Introduction of the Steam Engine into the Dutch Landscape". *Technology and Culture*, 45, pp. 251-276.

Roberts L. (2009). *Full steam ahead: Entrepreneurial engineers as go-betweens during the late eighteenth century*, in Schaffer S. *et al.* (ed.), *The Brokered World: Go-betweens and Global Intelligence, 1770-1820*. Camden: Science History Publications.

Roberts L. (2011). "Geographies of steam: mapping the entrepreneurial activities of steam engineers in France during the second half of the eighteenth century". *History and Technology*, 27, pp. 417-439.

Serraglio R. (2003). *L'Acqua Carolina per l'approvvigionamento idrico di insediamenti produttivi e centri urbani*, in Gambardella A. (ed.), *Napoli-Spagna: architettura e città nel XVIII secolo*. Naples: Edizioni Scientifiche Italiane.

Smiles S. (1865). *Lives of Boulton and Watt*. London: John Murray.

Tann J. (1978). "Marketing Methods in the International Steam Engine Market: The Case of Boulton and Watt". *The Journal of Economic History*, 38, pp. 363-391.

Temkin I. (2012). *At the Dawn of Malacology: The Salient and Silent Oeuvre of Giuseppe Saverio Poli*, in Baione T. (ed.), *Natural histories: extraordinary rare book selections from the American Museum of Natural History Library*. New York: Sterling.

Torrens H. (2006). *The geological work of Gregory Watt, his travels with William Maclure in Italy (1801–1802), and Watt's "proto-geological" map of Italy (1804)*, in Vai G.B. and Caldwell W.G.E. (eds.), *The Origins of Geology in Italy*. Boulder: The Geological Society of America.

Verbruggen J.A. (2005). "The correspondence of Jan Daniel Huichelbos van Liender with James Watt". Enschede: University of Twente.

Zorzanello G. (1984). "L'inedita corrispondenza del diplomatico veneziano Simon Cavalli con Matthew Boulton (1779-1786)". *Archivio Veneto*, CXXII, p. 35-64.